\def\ra{\rangle}
\def\la{\langle}
\def\t{\tilde}

\def\cb{{\Bbb C}}

\documentclass[fleqn,10pt]{wlscirep}

\usepackage{graphicx,color}
\usepackage{bbold}
\usepackage{graphicx, color, graphpap}
\usepackage{amsmath}
\usepackage{enumitem}
\usepackage{amssymb}
\usepackage{amsthm}
\usepackage{pstricks}
\usepackage{float}
\usepackage{multirow}
\usepackage{subfigure}
\usepackage{cite}
\long\def\ca#1\cb{} 

\newcommand{\Tr}{{\rm Tr}}

\renewcommand{\leq}{\leqslant}

\def\ra{\rangle}
\def\la{\langle}
\newcommand{\be}{\begin{equation}}
\newcommand{\ee}{\end{equation}}
\newcommand{\ba}{\begin{array}}
\newcommand{\ea}{\end{array}}

\newtheoremstyle{example}{\topsep}{\topsep}%
{}
{}
{\bfseries}
{.}
{   }
{\thmname{#1}\thmnumber{ #2}}
\theoremstyle{example}

\theoremstyle{definition}

\def\qed{\leavevmode\unskip\penalty9999 \hbox{}\nobreak\hfill
     \quad\hbox{\leavevmode  \hbox to.77778em{%
               \hfil\vrule   \vbox to.675em%
               {\hrule width.6em\vfil\hrule}\vrule\hfil}}
     \par\vskip3pt}

\title{Quantum steerability based on joint measurability}

\author[1]{Zhihua Chen}
\author[2,3]{Xiangjun Ye}
\author[4,5,*]{Shao-Ming Fei}
\affil[1]{Department of Mathematics, College of Science, Zhejiang University of
Technology, Hangzhou 310023, China}
\affil[2]{Key Laboratory of Quantum Information, University of Science and Technology of China, CAS, Hefei 230026, China}
\affil[3]{Synergetic Innovation Center of Quantum Information and Quantum Physics, University of Science and Technology of China, Hefei, 230026, China}
\affil[4]{School of Mathematical Sciences, Capital Normal University, Beijing 100048, China}
\affil[5]{Max-Planck-Institute for Mathematics in the Sciences, 04103 Leipzig, Germany}
\affil[*]{Corresponding-author: feishm@cnu.edu.cn}

\begin{abstract}

Occupying a position between entanglement and Bell nonlocality, Einstein-Podolsky-Rosen (EPR) steering has
attracted increasing attention in recent years. Many criteria have been proposed and experimentally implemented to characterize EPR-steering.
Nevertheless, only a few results are available to quantify  steerability using analytical results.
In this work, we propose a method for quantifying the steerability in two-qubit quantum states in the two-setting EPR-steering scenario,
using the connection between joint measurability and steerability.
We derive an analytical formula for the steerability of a class of X-states. The sufficient and necessary conditions for
two-setting EPR-steering are presented.
Based on these results, a class of asymmetric states, namely, one-way steerable states, are obtained.

\end{abstract}

\begin{document}
\flushbottom
\maketitle

\section*{Introduction}

Quantum nonlocality, EPR-steering and quantum entanglement are important quantum correlations. EPR-steering, which was originally presented by
Schrodinger in the context of the famous Einstein-Podolsky-Rosen (EPR) paradox \cite{Einstein}, lies between
quantum nonlocality and quantum entanglement,  which means that one observer,
by performing a local measurement on one's subsystem, can
nonlocally steer the state of the other subsystem. Recently EPR-steering was reformulated by Wiseman et al,
who described the hierarchy among Bell nonlocality, EPR-steering and quantum entanglement \cite{Wiseman}. EPR-steering has been shown to be advantageous for  quantum tasks such as randomness generation, subchannel discrimination, quantum information processing
and one-sided device-independent processing in quantum key distributions \cite{randomness, Piani, Bran, Bran2, Chen2}.

Many efforts have been made to detect and measure EPR-steering. Some steering inequalities based on uncertainty relations \cite{Reid, Cavalcanti, Schneeloch, Pramanik,  Schneeloch2, Kogias}, inequalities based on steering witnesses and the Clauser-Horne-Shimony-Holt (CHSH)-like inequality, and geometric Bell-like
inequalities et al\cite{Cavalcanti2, Roy, M, Semi, Saunders, Walborn, Ji}
are constructed to diagnose the steerability, are usually necessary
conditions. In addition to inequalities, all-versus-nothing proof without
inequalities,  were also presented to detect steerability \cite{Chen3}.
However only a few methods are available to quantify EPR-steering based on maximal violation of steering inequalities \cite{Hsieh}, steering
weight \cite{Skrzypczyk} and steering robustness. In these cases semi-definite programming is necessary to calculate the measures.
Recently, the radius of a super quantum hidden state model was proposed to evaluate the steerability \cite{Sun} by finding the optimal
super local hidden states. Nevertheless, it is formidably difficult to
find the optimal super quantum hidden states. A critical radius was proposed via the geometrical method, and the critical radius of T-states was calculated explicitly \cite{EPL}.
The closed formulas for steering were derived in two- and three-measurement scenarios \cite{Costa}, which is the case in which Alice and Bob are both allowed to measure the observables at their own sites.
It has been proven that one-to-one mapping exists between the joint measurability and the steerability of any assemblage \cite{map, Quintino, PRL113, Cavalvanti3}.
Using the connection between steering and joint measurability, the closed formula of the measure for two-setting EPR-steering of Bell-diagonal states was given \cite{Quan}. However, for any two-qubit quantum states, one still lacks a closed formula for the steerability problem, even for a 2-setting scenario.

Different from Bell nonlocality and quantum entanglement, steering exhibits asymmetric features, as proposed by Wiseman et al \cite{Wiseman}. There exist quantum states $\rho_{AB}$, for which
Alice can steer Bob's state but Bob can not steer Alice's state, or vice versa. This distinguishing feature could be useful for some one-way quantum information tasks such as quantum cryptography,
but until recently only a few asymmetric states have been proposed  and experimentally demonstrated \cite{Handchen, Bowles, Sun,Xiao}.

In this work, we investigate the analytical formula for quantification of EPR-steering  and obtain the necessary and sufficient condition of steerability for a class of quantum states. The asymmetric feature of EPR-steering is also investigated.

\subsection*{Setting up the stage}
Consider a bipartite qubit system $\rho_{AB}$ shared by Alice and Bob with reduced density states $\rho_A$ and $\rho_B$.
Alice performs positive-operator-valued measures (POVMs) $\Pi_{\kappa|\vec{n}}$ on subsystem $A,$ where
$\Pi_{\kappa|\vec{n}}=\frac{1}{2}(\emph{I}_2+(-1)^{\kappa} {\vec{n}}\cdot {\vec\sigma}),$ $\emph{I}_2$
is the identity
matrix and $\vec\sigma=(\sigma_x, \sigma_y, \sigma_z)$ are the Pauli matrices. Alice obtains the result $\kappa~(\kappa=0,1)$ when measuring
along the direction $\vec{n}.$ Bob's unnormalized conditional state is
$\tilde{\rho}_{\kappa|\vec{n}}=\Tr_A[\rho_{AB}(\Pi_{\kappa|\vec{n}}\otimes\emph{I}_2)]$.
 Bob's unconditional state
$\rho_B=\Tr_A\rho_{AB}=\sum\limits_{\kappa}\tilde{\rho}_{\kappa|\vec{n}}$ remains unchanged under any measurement direction.
A state assemblage ${\tilde{\rho}}_{\kappa|\vec{n}}$ is unsteerable if there exists a local hidden state model (LHSM)
with the state ensemble of ${p_i\rho_i}$ satisfying  $\tilde{\rho}_{\kappa|\vec{n}}=\sum\limits_i P(\kappa|\vec{n},i)p_i\rho_i$,
where $\rho_B=\sum\limits_i p_i\rho_i$ and $\sum\limits_{\kappa} P(\kappa|\vec{n},i)=1.$ The quantum state $\rho_{AB}$ is unsteerable
from $A$ to $B$ if for all local POVMs, the state assemblages are all unsteerable.
The quantum state $\rho_{AB}$ is steerable from $A$ to $B$ if there exist measurements in Alice's case that produce an assemblage
that demonstrates steerability.

The corresponding local hidden state model and the joint measurement observables are connected through $O_{\kappa|\vec{n}}=\frac{1}{\sqrt{\rho_B}}\tilde{\rho}_{\kappa,\vec{n}}\frac{1}{\sqrt{\rho_B}}$ and $G_i=\frac{1}{\sqrt{\rho_B}}p_i\rho_i\frac{1}{\sqrt{\rho_B}}$ by the one-to-one
mapping between the joint measurement problem and the steerability problem, whenever $\rho_B$ is invertible \cite{map}.
The steerability can be detected through the joint measurability of the observables.

Two-setting steering scenario: Any two-qubit quantum state can be expressed by $\rho_{AB}=(\emph{I}_4+\vec{a}\cdot\vec{\sigma}\otimes\emph{I}_2+\emph{I}_2\otimes \vec{b}\cdot\vec{\sigma}+\sum\limits_{i}^3c_{i}\sigma_i\otimes\sigma_i)/4$ under local unitary equivalence,
where ${\vec{a}, \vec{b}, \vec{c}}\in R^3$, $\sigma_1=\sigma_x$, $\sigma_2=\sigma_y$,
$\sigma_3=\sigma_z$, $\vec{\sigma}=\{\sigma_1, \sigma_2, \sigma_3\}$, $C=\rm{Diag}\{c_1, c_2, c_3\}$ is the correlation matrix.

When Alice performs two sets of POVMs $\Pi_{\kappa|\vec{n}_i}=(\emph{I}_2+(-1)^{\kappa} {\vec{n}_i}\cdot {\vec\sigma})/2$ $(i=0,1,\,\kappa=0,1)$ on $A$
with $\vec{n}_i=(\sin\alpha_i\cos\beta_i,\\
\sin\alpha_i\sin\beta_i, \cos\alpha_i),$
Bob's unnormalized conditional states are
$\tilde{\rho}_{\kappa|\vec{n}_i}={\Tr[\tilde{\rho}_{\kappa|\vec{n}_i}]}(\emph{I}_2+(-1)^{\kappa}\vec{s}_{\kappa,i}\cdot \vec{\sigma})/2$,
where $\Tr[\tilde{\rho}_{\kappa|\vec{n}_i}]=(1+(-1)^{\kappa}\vec{a}\cdot\vec{n}_i)/2$ and $\vec{s}_{\kappa,i}=(\vec{b} + (-1)^{\kappa} C\cdot\vec{n}_i)/(2\Tr[\tilde{\rho}_{\kappa|\vec{n}_i}])$.
When $|b|\neq 1,$ the measurement assemblages are
$$\begin{array}{rcl}
O_{\kappa}(x_i,\vec{g}_i)&=&\frac{1}{\sqrt{\rho_B}}\,\tilde{\rho}_{\kappa|\vec{n}_i}\,\frac{1}{\sqrt{\rho_B}}=\frac{1}{2}((1+(-1)^{\kappa}x_i)\emph{I}_2+(-1)^{\kappa}\vec{g}_i\cdot\vec{\sigma}),
\end{array}
$$
where $\vec{g}_i=U \,\vec{n}_i,$ $x_i=V\,\vec{n}_i$
with
$$
U=\frac{\vec{b}\,\vec{a}^{T}}{|b|^2-1}+\frac{(-1+\sqrt{1-|b|^2})\vec{b}\,\vec{b}^T C}{|b|^2(|b|^2-1)}+\frac{C}{\sqrt{1-|b|^2}},
$$
and $V=\frac{\vec{a}^{T}-\vec{b}^{T} C}{1-|b|^2}.$
Thus, $\{\tilde{\rho}_{\kappa|\vec{n}_i}\}_{\kappa, i}$  are unsteerable assemblages  if and only if
$\{O_{\kappa}(x_i,\vec{g}_i)\}_{\kappa, i}$  are jointly measurable \cite{unsharp, QIP, coex}, namely,
\begin{equation}\label{jm}
(1-F_{x_0}^2-F_{x_1}^2)(1-\frac{x_0^2}{F_{x_0}^2}-\frac{x_1^2}{F_{x_1}^2})-(\vec{g_0}\cdot\vec{g_1}-x_0 x_1)^2 \leq 0,
\end{equation}
where
$F_{x_i}=\frac{1}{2}(\sqrt{(1+x_i)^2-g_i^2}+\sqrt{(1-x_i)^2-g_i^2}),$ $g_i=|\vec{g}_i|.$

(\ref{jm}) gives rise to the condition for Alice to steer Bob's state. If Bob performs two sets of POVMs $\Pi_{\kappa|\vec{n}_i}$ on his system to
steer Alice's state, the corresponding condition can be similarly written by changing $\vec{a}\rightarrow \vec{b}$, $\vec{b}\rightarrow \vec{a}$ and
$C\rightarrow C^T$ in (\ref{jm}).

However, it is generally quite difficult to address condition (\ref{jm}) and obtain explicit conditions to judge the steerability for an arbitrary given
two-qubit state. For Bell-diagonal states, a necessary and sufficient condition of steerability has been derived from the relations between steerability and the joint measurability
problem \cite{Quan}. In the following, we study the steerability of any arbitrary given two-qubit states. We present analytical steerability conditions
for classes of two-qubit $X$-state.

\section*{Results}
\subsection*{Steerability of two-qubit states}

First, based on the jointly measurable condition (\ref{jm}) of
$\{O_\kappa(x_i,\vec{g}_i)\}_{\kappa, i}$
for the two-setting steering scenario, we define
the steerability of two-qubit states $\rho_{AB}$ by the following
\begin{equation}\label{S}
S =\max\{\max\limits_{\alpha_i,\beta_i} (S_1-S_2), 0\},
\end{equation}
where $S_1=(1-F_{x_0}^2-F_{x_1}^2)(1-\frac{x_0^2}{F_{x_0}^2}-\frac{x_1^2}{F_{x_1}^2})$, $S_2=(\vec{g}_0\cdot\vec{g}_1-x_0 x_1)^2$,
and the maximization runs over all of the measurements $\Pi_{\kappa|\vec{n}_i}$, namely, over the parameters $\alpha_i$ and $\beta_i$, $i=0,1$.
It is obvious that $S$ lies between $0$ and $1,$ and $\rho_{AB}$ is steerable if and only if $S>0$.

For general two-qubit states, a global search can be used to obtain the global minimum values of $S.$
The Matlab code is supplied in the supplementary material.

Due to the relationship between the joint measurements and steerability, local hidden states $\tilde{\rho}_{\kappa|\vec{n}_i}$ are
represented as $\sqrt{\rho_B}G_{\mu v}\sqrt{\rho_B}$ $(\mu=\pm 1, v=\pm 1),$  where $G_{\mu v}=\frac{1}{4}(1+\mu x_0 + v x_1+\mu v Z+(\mu v \vec{z}+\mu \vec{g}_0 + v \vec{g}_1)\vec{\sigma})$ which are all possible sets of four measurements  satisfying the marginal constraints for
any two jointly measurable observables $\{O_\kappa(x_i,\vec{g}_i)\}_{\kappa, i}$ \cite{unsharp, QIP, coex}.
The steering radius $R(\rho_{AB})$ \cite{Sun} can be calculated by optimizing $\vec{z}$ and $Z.$

In the following, we analytically calculate  the steerability $S$ for some $X$-states $\rho_X$.
We define a class of two-qubit X-states to be zero-states $\rho_{zero}$ if the $X$-states $\rho_X$ satisfy
the condition that the maximum points (stationary points)  of $S_1$ belong to the zero points of $S_2$
with respect to the measurement parameters $\alpha_i$ and $\beta_i, (i=1,2)$.

For any two-qubit X-state,
$\rho_X=\frac{1}{4}(\emph{I}_4+ a_3 \sigma_3\otimes\emph{I}_2+ b_3 \emph{I}_2\otimes \sigma_3+\sum\limits_{i}^3 c_{i} \sigma_i\otimes\sigma_i)$, we have
$U=\rm{Diag}\{u_1,u_2,u_3\},$ $V=[0,0,t_3]$, where $u_1={c_{1}}/{\sqrt{1-b_3^2}}$, $u_2={c_{2}}/{\sqrt{1-b_3^2}}$, $u_3=({a_3 b_3-c_{3}})/({-1+b_3^2})$ and $t_3=({a_3-b_3 c_3})/({1-b_3^2}).$
We obtain the following results:

{\bf{Theorem.}}~ For the zero-states $\rho_{zero}$, the analytical formula of the steerability is given by
\begin{eqnarray}\label{AS}
S=\max\{\Delta_1, \Delta_2, \Delta_3, 0\},
\end{eqnarray}
where $\Delta_1=u_1^2+u_2^2-1,$ $\Delta_2=\frac{1}{2}[u_1^2(u_3^2-t_3^2)+u_1^2+u_3^2+t_3^2-1-(1-u_1^2)\sqrt{((1-t_3)^2-u_3^2)((1+t_3)^2-u_3^2)}],$
$\Delta_3=\frac{1}{2}[u_2^2(u_3^2-t_3^2)+u_2^2+u_3^2+t_3^2-1-(1-u_2^2)\times\sqrt{((1-t_3)^2-u_3^2)((1+t_3)^2-u_3^2)}].$
When $S>0,$ the optimal measurements that give rise to maximal $S$ are $\sigma_x$ and $\sigma_y$
if $\Delta_1>\max\{\Delta_2, \Delta_3, 0\},$ $\sigma_x$ and $\sigma_z$ if $\Delta_2>\max\{\Delta_1, \Delta_3, 0\},$
and $\sigma_y$ and $\sigma_z$ if $\Delta_3>\max\{\Delta_1, \Delta_2, 0\}$.

The proof is given in the supplementary material.

It is obvious that any X-state with $t_3=0$  belongs to $\rho_{zero}$, e.g.,
$|\phi\rangle = a |00\rangle +\sqrt{1-a^2} |11\rangle$ $(0<|a|<1)$ and the Bell-diagonal state
$\rho =\frac{1}{4}(\emph{I}+ c_1 \sigma_1 \otimes \sigma_1 + c_2 \sigma_2 \otimes \sigma_2 + c_3 \sigma_3 \otimes \sigma_3)$ are all the zero states.
For $|\phi\rangle,$ we have $S=1$.

For the Bell-diagonal state, interestingly, the steerability $S$ is given by the non-locality
characterized by the maximal violation of the CHSH inequality.
Let ${\mathcal{B}_{CHSH}}$ denote the Bell operator for the CHSH inequality \cite{chsh},
${\mathcal{B}_{CHSH}}=A_1\otimes B_1+A_1\otimes B_2+A_2\otimes B_1-A_2\otimes B_2$,
where $A_i=\vec{a}_i \cdot \vec{\sigma}$, $B_i=\vec{b}_i \cdot \vec{\sigma}$, $\vec{a}_i$ and $\vec{b}_i$, $i=1,2$,
are unit vectors. Thus, the the maximal violation of the CHSH inequality is given by \cite{ho340}
\be\label{maxv}
N=\max_{\mathcal{B_{CHSH}}}|\la{\mathcal{B_{CHSH}}}\ra_{\rho}|=2\sqrt{\tau_1+\tau_2},
\ee
where $\tau_1$ and $\tau_2$ are the two largest eigenvalues of the matrix $T^{\dag}T$, $T$ is the matrix with
entries $T_{\alpha\beta}=tr[\rho\,\sigma_\alpha\otimes\sigma_\beta]$, $\alpha,\,\beta=1,2,3$, $\dag$ indicates transpose and conjugation.
For the Bell-diagonal state, we have
$N=2\sqrt{c_1^2+c_2^2+c_3^2-\min\{c_1^2,c_2^2,c_3^2\}}$.
From (\ref{AS}), we find that the steerability of Bell-diagonal state is given by $S=\frac{N^2}{4}-1$.

For $t_3\neq 0$, we give the explicit conditions of the zero states in the supplementary material.

In the following, we present the maximum value of the steerability $S$ for a given $N$ of $\rho_{zero}$.

{\bf{Corollary 1:}} For zero-states $\rho_{zero}$ with given $N$, $0\leq N\leq 2$, we have
$S\leq \frac{N}{2}$. Moreover, $S={N}/{2}$ is attained when $a_3 = 1-c_3+b_3,$  $b_3 \rightarrow -1,$ $c_{1}=\sqrt{(1+b_3)(c_3-b_3)},$ $c_{2}=-c_{1},$
i.e., $\rho_{zero}$ has the following form,
\be
\rho_{X_0}=\left(\begin{matrix}
\frac{1+b_3}{2} & 0 & 0 & \pm\frac{\sqrt{(1+b_3)(c_3-b_3)}}{2}\\
0 & \frac{1-c_3}{2} & 0 &0\\
0 & 0 & 0 &0\\
\pm\frac{\sqrt{(1+b_3)(c_3-b_3)}}{2} & 0 & 0 &\frac{c_3-b_3}{2}\end{matrix} \right).\label{upper}
\ee

The following corollary gives the conditions at which we obtain the minimal value of $S$ for a given $N$.

{\bf{Corollary 2:}}
For zero-states $\rho_{zero}$ with given CHSH value $N$, $S$ obtains the minimal value
when $a_3=0$ and $b_3=0$ or $|a_3+b_3|=\sqrt{(1+c_3)^2-(c_1-c_2)^2}$ or $|a_3-b_3|=\sqrt{(1-c_3)^2-(c_1+c_2)^2}$.

The proofs of Corollary 1 and Corollary 2 are given in  the supplementary material.
In Fig. 1, we give a description for the boundaries of the steerability $S$ for a given value of $N$.
From Fig. 1, we observe that for any given $N$ with $0 \leq N \leq 2$, the lower bound of $S$ is always 0 and the upper bound of $S$ is always
less than 2 (light blue), and for $N > 2,$ the lower bound of $S$ is always greater than 0, and the upper bound of $S$
is always $2$ (dark blue).

\begin{figure}[ht]
\centering
\includegraphics[width=0.3\linewidth]{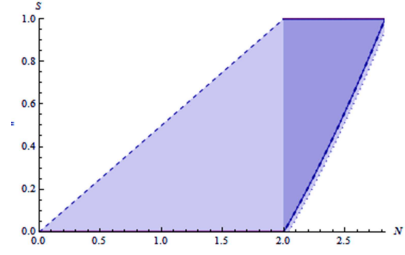}
\caption{Regions of the values taken on by steerability $S$ for given $N$.}
\label{fig:Fig1}
\end{figure}

For zero-states $\rho_{zero},$ the steering radius $R(\rho_{zero})$  can be obtained when Alice measures her
qubit along the directions $\sigma_x$ and $\sigma_y,$  or $\sigma_x$ and $\sigma_z,$ or $\sigma_y$ and $\sigma_z.$
Indeed, from the construction of joint measurements \cite{unsharp}, when Alice measures her
qubit along the directions of $\sigma_x$ and $\sigma_z,$ the local hidden states can be expressed as follows
$$
\frac{1}{2}(\emph{I}_2+\frac{m_x \sigma_x+m_z \sigma_z}{1+\mu a_3+v(b_3 z_3+Z)}),
$$
where $m_x=\mu v(c_1+\mu\sqrt{1-b_3^2}z_1),$ $m_z=b_3+\mu c_3 +v(z_3 + b_3 Z),$ $\mu=\pm 1, v=\pm 1.$
Therefore
\begin{equation}
R(\rho_{zero})=\max\{r(\rho_x)_{xy},r(\rho_x)_{xz},r(\rho_x)_{yz}\},
\end{equation}
where
\begin{equation}
\begin{aligned}
&r(\rho_{zero})_{xy}=\sqrt{c_1^2+c_2^2+b_3^2}; \hspace{1cm}r(\rho_{zero})_{xz}=\min\limits_{z_1,z_3,Z}\max\limits_{\mu,v}\sqrt{r^{xz}_{\mu,v}}; \hspace{1cm}r(\rho_{zero})_{yz}=\min\limits_{z_1,z_3,Z}\max\limits_{\mu,v}\sqrt{r^{yz}_{\mu,v}};\\ \nonumber
&r^{xz}_{\mu,v}=\frac{(c_1+\mu\sqrt{1-b_3^2}z_1)^2+(b_3+\mu c_3+v(z_3+b_3Z))^2}{(1+\mu a_3+v(b_3z_3+Z))^2};\hspace{1cm}r^{yz}_{\mu,v}=\frac{(c_2+\mu\sqrt{1-b_3^2}z_1)^2+(b_3+\mu c_3+v(z_3+b_3Z))^2}{(1+\mu a_3+v(b_3z_3+Z))^2}. \nonumber
\end{aligned}
\end{equation}
It is not easy to calculate $r(\rho_{zero})_{xz}$ and $r(\rho_{zero})_{yz}$ analytically.
We give the analytical results for $R(\rho_{zero})$ for some special states in the following.

\subsection*{Asymmetric two-setting EPR-steering}
Different from Bell-nonlocality and quantum entanglement, EPR-steering has the asymmetric property of
one-way EPR steering: Alice may steer Bob's state but not vice versa.
The demonstration of asymmetric steerability has practical implications in quantum communication networks \cite{Wollmann}.
Until now, only a few asymmetric steering states have been found \cite{Handchen, Bowles, Sun,Xiao}.
In this work we present a class of asymmetric steering states of the form $\rho_{X_0}$ in (\ref{upper}).

If Alice performs measurements on her qubit,
the steerability is given by $S(\rho_{X_0})=\max\{\frac{2c_3-1-b_3}{1-b_3},0\}$ which approaches $c_3$ when $b_3$ approaches $-1$ and $c_3>0$.
If Bob performs measurements on his qubit, the related steerability is given by the following
$$
S(\rho_{X_0})=\max\{\frac{(1+b_3)(b_3+c_3)}{(2+b_3-c_3)^2},0\}
$$
which is equal to zero as long as $(1+b_3)(b_3+c_3)\leq 0$.
Therefore, when $0<c_3<-b_3$ and $b_3\rightarrow-1,$ Alice can always steer Bob's state, but Bob can never steer Alice's state
(see Fig. 2 for the asymmetric EPR-steering for $b_3=-0.999$). We note that Alice can always steer Bob's state,
but Bob can not steer Alice's state.

\begin{figure}[ht]
\centering
\includegraphics[width=0.3\linewidth]{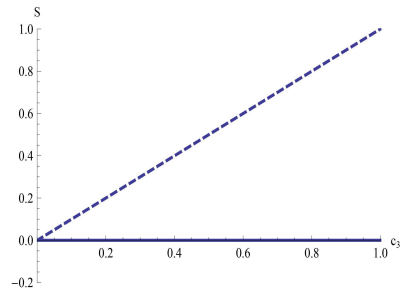}
\caption{Steerability $S$ versus $c_3$ for $b_3=-0.999$.
The dashed line indicates Alice steering Bob's state, and the solid line (horizontal coordinate) denotes Bob steering Alice's state.}
\label{fig:Fig2}
\end{figure}

In the following subsection, we investigate the geometric features of the asymmetric steering state-$\rho_{x_0}$ in terms of the steering ellipsoid \cite{Ellipsoid}.
The steering ellipsoid of $\rho_{X_0}$ when Alice performs POVMs is quite different from that  when Bob performs POVMs.
The centre of the steering ellipsoid $\varepsilon_B$ for Alice performing POVMs on her qubit is $(0,0,({b_3-a_3 c_3})/({1-a_3^2}))$,
which goes to $(0,0,-1)$ when $b\rightarrow -1,$ and the volume of the steering ellipsoid $\varepsilon_B$ is given as follows
$$\frac{4\pi}{3}\frac{|c_1c_2(c_3-a_3b_3)|}{(1-a_3^2)^2}=\frac{4\pi}{3}\frac{(1+b_3)^2}{(2-c_3+b_3)^2},$$
In this case the steering ellipsoid is tangent to the Bloch sphere.
The centre of the steering ellipsoid $\varepsilon_A$ for Bob performing POVMs on his qubit is
$$
(0,0,\frac{a_3-b_3 c_3}{1-b_3^2}=(0,0,\frac{1-c_3}{1-b_3}),
$$
which goes to ${(1-c_3)}/{2}$ when $b_3\rightarrow -1$.
The volume of the steering ellipsoid $\varepsilon_A$ is given by the following
$$
\frac{4\pi}{3}\frac{|c_1c_2(c_3-a_3b_3)|}{(1-b_3^2)^2}=\frac{4\pi(c_3-b_3)^2}{3(1-b_3)^2},
$$
which goes to $\frac{\pi(1+c_3)^2}{3}$ when $b_3\rightarrow -1$.
The steering ellipsoid is also tangent to the Bloch sphere.
In this case the ellipsoid shows some peculiar features, i.e., when $b_3\rightarrow -1$ and $c_3\rightarrow 0$, the ellipsoid $\varepsilon_B$ is nearly $0$,
but Alice can still steer Bob; however, when $b_3\rightarrow -1$ and $c_3\rightarrow -b_3$, the ellipsoid $\varepsilon_A$ is almost the entire Bloch sphere,
but Bob can not steer Alice.

As a special case of $\rho_{X_0},$ we take $a_3=1-2\eta(1-\chi),$ $b_3=2\eta\chi-1,$ $c_3=2\eta-1,$ $c_1=-c_2=-2\eta\sqrt{\chi(1-\chi)}$.
The state has the following form,
\be
W_\eta^\chi=\left(\begin{matrix}
\eta\chi & 0 & 0 & -\eta\sqrt{\chi(1-\chi)}\\
0 & 1-\eta & 0 &0\\
0 & 0 & 0  &0\\
-\eta\sqrt{\chi(1-\chi)} & 0 & 0 &\eta(1-\chi)\end{matrix} \right).\label{wiseman}
\ee
From the theorem, we obtain the following when Alice measures her qubit,
\begin{equation}
\begin{aligned}
S(W_\eta^\chi)=&\max\{\frac{1+\eta(-2+\chi)}{-1+\eta\chi},\frac{\eta(1+\eta(-2+\chi))(-1+\chi)}{(1-\eta\chi)^2},0\}\nonumber.
\end{aligned}
\end{equation}
The sufficient and necessary condition in the two-setting steering scenario is
$\eta>{1}/({2-\chi})$ for Alice to steer Bob's state.
The corresponding optimal measurements are $\sigma_x$ and $\sigma_y.$

If Bob measures his qubit, the steerability is given by the following
$$
S(W_\eta^\chi)=\max\{\frac{\eta\chi(-1+\eta+\eta\chi)}{(1+\eta(-1+\chi))^2},\frac{-1+\eta+\eta\chi}{1+\eta(-1+\chi)}, 0\}.
$$
The sufficient and necessary condition for Bob to steer Alice's state is
$\eta>{1}/({1+\chi})$.
The related optimal measurements are
$\sigma_x$ and $\sigma_y.$
The asymmetric property in quantum steering given by this example is shown in Fig. 3 and Fig. 4.
The steering radius is $\sqrt{1-4\eta\chi(1-\eta(2-\chi))}$ when Alice measures her qubit,
and $\sqrt{1-4\eta(1-\chi)(1-\eta-\eta\chi)}$ when Bob measures his qubit.

\begin{figure}[ht]
\centering
\includegraphics[width=0.3\linewidth]{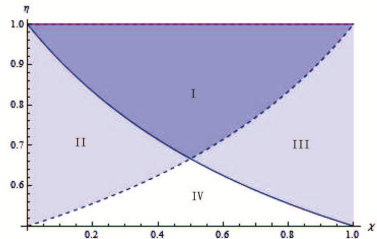}
\caption{Parameter region for which Alice (Bob) can steer Bob's (Alice's) state for the state $W_\eta^\chi$.
In region I, Alice can steer Bob's state, and Bob can also steer Alice's state. In region II (III),
Alice (Bob) can steer Bob's (Alice's) state, but Bob (Alice) can not steer Alice's (Bob's) state.
In region IV, Alice can not steer Bob's state, and Bob can not steer Alice's state.}
\label{fig:Fig3}
\end{figure}

\begin{figure}[ht]
\centering
\includegraphics[width=0.8\linewidth]{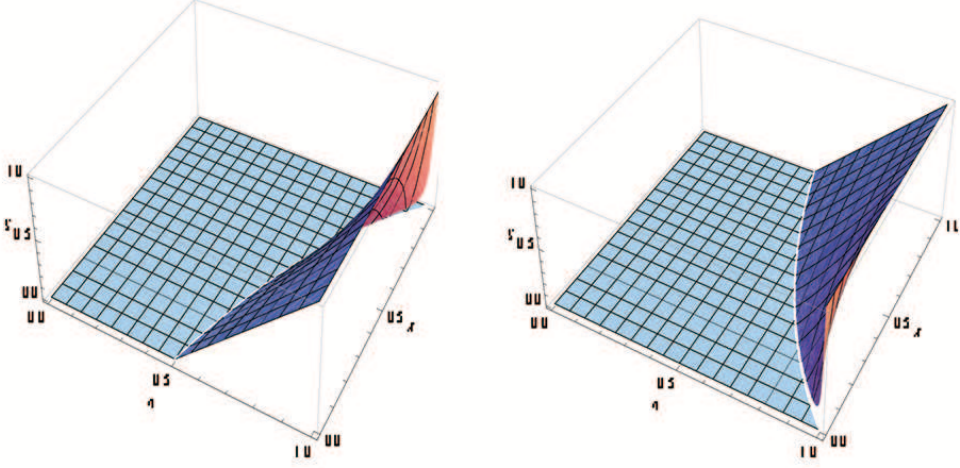}
\caption{The left figure(the right figure): $S(W_\eta^\chi)$ when Alice (Bob) measures her (his) qubit.}
\label{fig:Fig4}
\end{figure}

As another example showing the asymmetry of quantum steering, we consider the state $W_V^{\theta}$ \cite{Sun},
\begin{equation}
W_V^\theta=V |\psi_1\rangle\langle\psi_1|+(1-V)|\psi_2\rangle\langle\psi_2|,
\end{equation}
where $|\psi_1\rangle=\cos\theta |00\rangle+\sin\theta |11\rangle,$
$|\psi_2\rangle=\cos\theta |10\rangle+\sin\theta |01\rangle,$
$\theta\in(0,\pi/2),\;V\in[0,1/2)\cup(1/2,1]$.
$W_V^\theta$ is a zero state. From our theorem,
we know that when Alice performs measurements on her qubit, $S(W_V^\theta)=(1-2 V)^2$. The optimal measurements are $\sigma_x,$ $\sigma_y$ or $\sigma_x,$ $\sigma_z.$
This state is always steerable for Alice except when $V=1/2$.

When Bob performs two projective measurements on his qubit, we have the following
\begin{equation}
\begin{aligned}
S(W_V^\theta)=&\max\{\frac{(1-2 V)^2-\cos^22\theta}{1-(1-2 V)^2\cos^22\theta}, \frac{\sin2\theta^2((1-2 V)^2-\cos^22\theta)}{(1-(1-2 V)^2\cos^22\theta)^2},0\}.
\end{aligned}
\end{equation}
The sufficient and necessary condition in the two-setting steering scenario for Bob to steer Alice's state is $|\cos 2\theta|<|2 V-1|$,
with the optimal measurements $\sigma_x$ and $\sigma_y$. For $W_V^\theta,$ the corresponding steering radius is $\sqrt{1+(1-2 V)^2\sin^22\theta}$ when Alice measures her qubit,
and $\sqrt{(1-2 V)^2+\sin^22\theta}$ when Bob measures his qubit.
From Fig. 5 we observe that Alice can always steer Bob's state except when $V=1/2,$
but Bob can steer Alice's state only for some $V$ depending on $\theta$.

\begin{figure}[ht]
\centering
\includegraphics[width=0.3\linewidth]{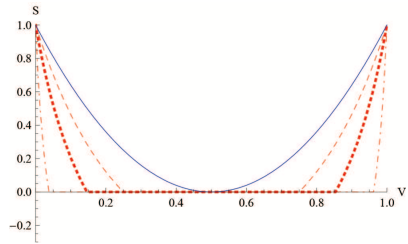}
\caption{$S(W_V^\theta)$ versus $\theta$: the blue solid line denotes when Alice measures her qubit, and the
red dashed line ($\theta=\frac{\pi}{6}$), red dotted line ($\theta=\frac{\pi}{8}$), and red dot-dashed line ($\theta=\frac{\pi}{16}$) indicate when Bob measures his qubit.}\label{fig5}
\end{figure}

From our theorem, the analytical results of steerability can be obtained for more detailed zero states, and the asymmetric property of steering
can be readily studied. In the following, we give two examples of symmetric two-setting EPR-steering.

{\it Example 1.} The two-qubit nonmaximally entangled state mixed with colour noise,
$$
\rho_{\rm cn}=V|\psi(\theta)\rangle\langle\psi(\theta)|+\frac{1-V}{2}(|00\rangle\langle 00|+|11\rangle\langle 11|),
$$
where $|\psi(\theta)\rangle=\cos\theta|00\rangle+\sin\theta|11\rangle$, $\theta\in(0,\pi/2)$, $V\in(0,1]$.
The steerability is given by $S(\rho_{\rm cn})={V^2\sin^22\theta}/({1-V^2\cos2\theta^2})$. Therefore,
$\rho_{\rm cn}$ is steerable if and only if $V\sin2\theta\neq 0.$



{\it Example 2.} The generalized isotropic state,
$\rho_{gi}=V|\psi(\theta)\rangle\langle\psi(\theta)|+(1-V)\emph{I}/{4}$,
where
$|\psi(\theta)\rangle=\cos\theta|00\rangle+\sin\theta|11\rangle$, $\theta\in(0,\pi/2)$, $V\in(0,1]$. The
state reduces to the usual isotropic state when $\theta=\pi/4$.
According to our theorem, we obtain the analytical steerability of $\rho_{gi}$,
\begin{eqnarray}\nonumber
S(\rho_{gi})=&\frac{1-V^2\cos^24\theta+(1-V)\sqrt{(1+V)^2-4V^2\cos^22\theta}}{4(1-V^2\cos^22\theta)}\times\frac{V^2(1+2\sin^22\theta)-1-(1-V)\sqrt{(1+V)^2-4V^2\cos^22\theta}}{1-V^2\cos^22\theta}.
\end{eqnarray}
Hence, the sufficient and
necessary condition of steerability is $1+(1-V)\sqrt{(1+V)^2-4V^2\cos^22\theta}<V^2(1+2\sin^22\theta).$

\section*{Discussion}
Based on the one-to-one correspondence between EPR-steering and joint measurability,
we have investigated the steerability for any two-qubit system in the two-setting measurement scenario.
The steerability we introduced is invariant under local unitary operations.
The analytical formula for steerability has been derived for a class of X-states, and
the sufficient and necessary conditions for two-setting EPR-steering have been presented.
For general two-qubit states, it has been shown that the lower and upper bounds of steerability are explicitly connected
to the non-locality of the states given by the CHSH values of maximal violation.
Moreover, we have also presented a class of asymmetric steering states
by investigating steerability with respect to the measurements from Alice's and Bob's sides.
Our strategy might also be used to study the quantification of steerability for multi-setting scenarios, in particular,
for three-setting scenarios for which the joint measurability problem of three qubit observables has
already been investigated \cite{three, JPA}. Our method might also be used in continuous variable steering, temporal and channel
steering, for which the steerability of the state assemblages or the instrument assemblages can be connected to the incompatibility problems of the quantum measurement assemblages \cite{CVsteering, unified}.
Hence, the steerability of the quantum states or the quantum channels might also be studied based on
the corresponding measurement incompatibility problems.

\bibliographystyle{plain}

\section{Acknowledgements}
This work is supported by the NSFC under no. 11571313,11475089, 11675113.

\section{Author contributions}
Z.C. and X.Y initiated the research, Z.C. proved the main theorems and developed the numerical codes, and Z.C. X.Y. and S.F. wrote the manuscript.

\section{Additional information}
\textbf{Competing financial interests:} The authors declare no competing financial interests.

\end{document}